# The Spectral Density of a Spin System Calculated for Solvable Models

B.V. Kryzhanovsky

Center of Optical Neural Technologies, Scientific Research Institute for System Analysis of the Russian Academy of Sciences, Moscow
kryzhanov@niisi.ras.ru

**Abstract.** The relationship between the spectral density and free energy of a spin system is considered. The analytical expressions allowing for the calculation of the spectral density for solvable models are determined. A linear Ising model is taken for testing the approach. Additionally, the approach is used to analyze the spectral density of a 2D Ising model, a Bethe-lattice model and a mean-field model. Even a small change of the spectral density is shown to be able to radically change the parameters of the system.

## 1. Introduction

The spectral density is a fundamental characteristic determining the behavior of a system. The knowledge of the spectral density allows us to easily calculate the partition function and then use it to derive all other parameters of the system. There are currently a few models for which exact solutions for free energies are known [1, 2]. Note that these solutions have been found without knowledge of the system spectral density, using only the most general considerations, matrix algebra [3, 4] or combinatorial methods [5]. At the same time, the expression of spectral density is so far known for only a single-dimensional Ising model [6]. This fact encouraged a few researchers (e.g. [7, 8]) to try to get expressions for the spectral density within the context of different approximations. Our study develops an approach for finding the spectral density for, at least, the systems with known free energy.

## 2. General expressions

Let us consider a spin system defined by the Hamiltonian:

$$E = -\frac{1}{2N}\sum_{i,j=1}^{N} J_{ij} s_i s_j, \qquad s_i = \pm 1 \tag{1}$$

where $N$ is the number of spins, $E$ is the energy per a single spin, $J_{ij}$ is a symmetric zero-diagonal matrix ($J_{ii} = 0$); the energy of the ground state will be denoted as $E_0$.

Let us demonstrate how we can determine the spectral density of the spin system when we know the expression for the free energy. The partition function has the form:

$$Z = \sum_E D(E) e^{-N\beta E} \tag{2}$$

where $\beta = 1/kT$ is the inverse temperature, and $D(E)$ is the spectral density, i.e. the degeneracy of the energy level. Let us represent the spectral density as

$$D(E) = e^{N\Psi(E)} \qquad (3)$$

and substitute it into (2). Then replacing summation with integration in (2), we can, with the accuracy to an irrelevant constant, represent the partition function in the form:

$$Z \sim \int_{-\infty}^{\infty} e^{N[\Psi(E)-\beta E]} dE \qquad (4)$$

Evaluating integral (4) by the saddle-point method, we get $Z \sim \exp[-Nf(\beta)]$, where

$$f(\beta) = \beta U - \Psi(U), \quad \frac{d\Psi(U)}{dU} = \beta \qquad (5)$$

The first of the expressions (5) defines the free energy, and the second implicitly defines the saddle point $E = U$, at which $d[\Psi(E) - \beta E]/dE = 0$. Actually, the quantity $U$ is no other than the internal energy, which we will make sure of below.

So, we have the expressions (5), which hold the sought-for spectral function $\Psi = \Psi(E)$, to determine the free energy. Simultaneous solution of equations (5) does not seem possible. However, it can be done by going over to another set of equations. Having differentiated the first of the equations (5), we obtain another set of equations:

$$\Psi(U) = \beta U - f(\beta), \quad U = \frac{df(\beta)}{d\beta} \qquad (6)$$

These equations relate $\Psi(U)$ and internal energy $U$ to the temperature $\beta$. It follows from the equations that when $\beta$ varies from $\beta = 0$ to $\beta = \infty$, $U$ changes from 0 to $E_0$, and for each $\beta$ we have a pair of values $U$ and $\Psi(U)$. By doing so we define function $\Psi(U)$. What is said above means that we can rewrite these expressions for any $E$ without binding ourselves to such notions as internal energy and temperature:

$$\Psi(E) = bE - f(b), \quad E = \frac{df(b)}{db} \qquad (7)$$

$$\frac{d\Psi}{dE} = b, \quad \frac{d^2\Psi}{dE^2} = \left(\frac{d^2f}{db^2}\right)^{-1} \qquad (8)$$

Here we introduced a new general variable $b$ to emphasize that equations (7) carry no temperature dependency. Besides, we added the expressions obtained from (7) for the derivatives of the spectral function $\Psi = \Psi(E)$. Below we will use these expressions in the analysis of the spectral density. Note that the expressions (6) are true only when $U \leq 0$. Correspondingly, formulae (7) – (8) define spectral function $\Psi(E)$ only for $E \leq 0$. Nevertheless, we can build function $\Psi(E)$ for $E > 0$ because the spectrum of the models under consideration is symmetrical $\Psi(E) = \Psi(-E)$.

Set of equation (7) is easy to solve numerically: knowing function $f(b)$ for each $b$ from (7), it is possible to get the value of $E$ and the corresponding value of the spectral function $\Psi(E)$, and for each $b$ from (8) the corresponding values of its derivatives. As we will see below, for some models it is possible to derive functions $b = b(E)$ from equation $E = df(b)/db$, and present $\Psi(E)$ in an explicit form.

**2. Examination of solvable models**

To examine the characteristics of the spectral density, we use well-known models that only allow for interactions between nearest neighbors and can be solved exactly. To make the formulae look simpler, we set the exchange constant equal to one: $J = 1$. Then the ground-state energy takes the form of $E_0 = -\frac{1}{2}q$, where $q$ is the number of neighbors.

2.1. Let us take a linear Ising model ($q = 2$) as an example to verify the correctness of the expressions we derived earlier. Function $f(b)$ for this model has the form:

$$f(b) = -\left[b + \ln\left(1 + e^{-2b}\right)\right] \tag{9}$$

Solving the second equation in (7) for $b$, we get $b = \frac{1}{2}\ln[(1-E)/(1+E)]$. Substituting the solution in (7), we obtain the well-known expression [6] for the spectral function:

$$\Psi(E) = -\frac{1-E}{2}\ln\left(\frac{1-E}{2}\right) - \frac{1+E}{2}\ln\left(\frac{1+E}{2}\right) \tag{10}$$

2.2. Let us consider 2D Ising model ($q = 4$). As Onsager's solution [9] is well known, we do not present it here.

Let us first consider particular limiting cases assuming for simplicity that $J = J' = 1$, where $J$ and $J'$ are coupling strengths along different lattice axes. When $b \to 0$, function $f(b)$ takes the form $f(b) = -\ln 2 - b^2$. Substituting this into (7), we find that $E = -2b$ and, correspondingly, the spectral distribution center is

$$\Psi(E) = \ln 2 - \frac{1}{4}E^2 \tag{11}$$

Letting $b \to \infty$, we get $f(b) = 2b - 10e^{-8b}$. Substituting this in (7), we get the expressions for the spectral function at the wings of the distribution. With $E \to E_0$ we have

$$\Psi(E) = \frac{1}{8}(E - E_0)\left[1 + \ln\frac{80}{E - E_0}\right], \tag{12}$$

When $E \to |E_0|$, the similar expression can be obtained using the property of symmetry $\Psi(E) = \Psi(-E)$.

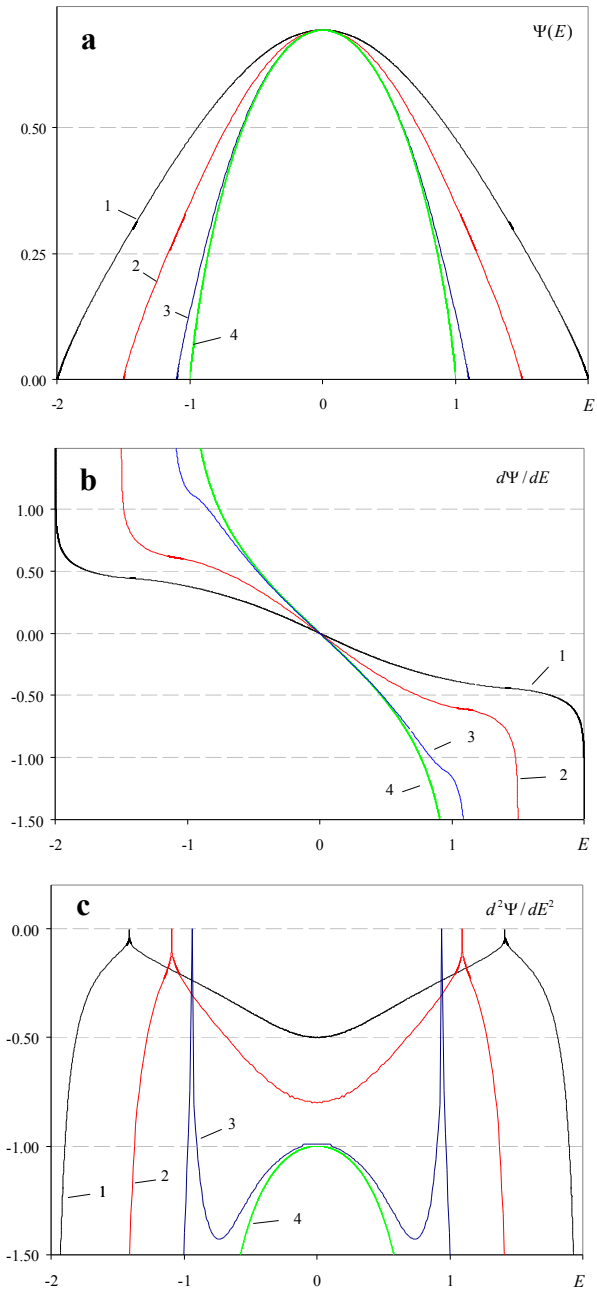
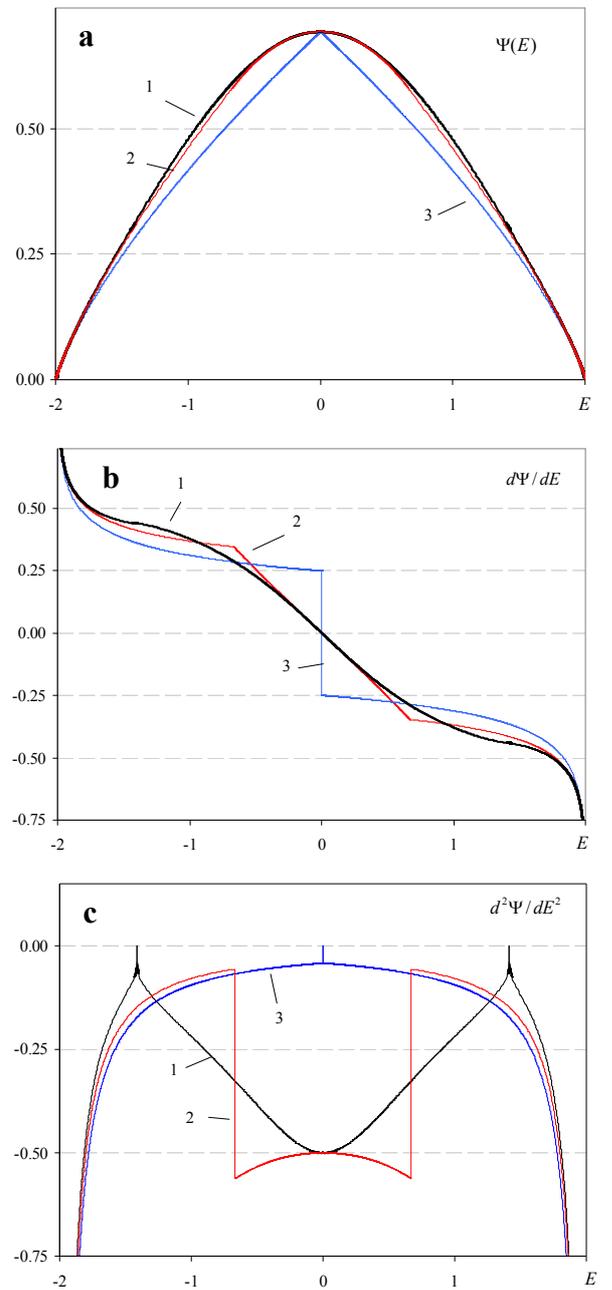

Fig. 1. Graphs of functions $\Psi(E)$, $d\Psi(E)/dE$ and $d^2\Psi(E)/dE^2$ for a 2D Ising model. Curves 1, 2, 3, 4 correspond to $J' = 1, 0.5, 0.1, 0$ at $J = 1$.

Fig. 2. Graphs of functions $\Psi(E)$, $d\Psi(E)/dE$ and $d^2\Psi(E)/dE^2$ at $q = 4$. Curves 1 is a 2D Ising model ($J' = J = 1$), curves 2 – the Bethe model, curves 3 – the mean-field model.

Unfortunately, the general analytical expression for $\Psi(E)$ is not possible to find. For this reason we will confine ourselves to graphical analysis. Figure 1a shows how the spectral function changes when coupling constant $J'$ varies from 0 to 1 ($J = 1$). As expected, when $J' \to 0$, the spectral function for the selected model coincides with the curve (10) corresponding to a 1D Ising model. At the same time, it is hard to use the curves of function $\Psi(E)$ to judge whether function $f(b)$ has a singularity. It is more convenient to use the curves of the derivatives of $\Psi(E)$ for this purpose. As we can see (Figures 1b and 1c), the first derivative has an inflection (the convexity passes into concavity) at a certain critical value $E = E_c$ ($E_c = E_0/\sqrt{2}$ at $J = J' = 1$), and the second derivative turns into zero. This fact signifies that the function $f(b)$ has a singularity: from (8) it follows that $d^2 f / db^2 \to \infty$ if $E \to E_c$. The energy interval near $E_c$ that provides the presence of singularity is very narrow: the smooth curve of the second derivative has small sharp peaks. As might be expected, these peaks and the singularity are gone when $J' \to 0$.

2.3. Let us consider a Bethe-lattice-based Ising model [10]. For this model we have:

$$2f(b) = -q[b + \ln(1 - z^2)] + \ln[1 + z^2 - z(x + x^{-1})] + (q - 2)\cdot\ln[x + x^{-1} - 2z] \quad (13)$$

where $z = e^{-2b}$, and the parameter $x$ is equal to the least of the positive solutions of the equation

$$z = x\frac{1 - x^{q-2}}{1 - x^q} \quad (14)$$

When $b \leq \tfrac{1}{2}\ln[q/(q-2)]$, the solution of (14) is $x = 1$ and the equation $E = df(b)/db$ can be solved for $b$ explicitly:

$$b = \frac{1}{2}\ln\left(\frac{1-\varepsilon}{1+\varepsilon}\right), \quad \varepsilon = E/E_0 \quad (15)$$

Correspondingly, from (7) we get the explicit form of the spectral function:

$$\Psi(E) = \ln 2 - \frac{q}{4}[(1-\varepsilon)\ln(1-\varepsilon) + (1+\varepsilon)\ln(1+\varepsilon)] \quad (16)$$

which is true for $|E| \leq |E_0|/(q-2)$. It is easy to see that on this energy interval the form of the spectral function (16) is similar to the expression (10) for a 1D Ising model and transforms into it when $q = 2$.

On the energy interval $|E_0| \geq |E| \geq |E_0|/(q-2)$, which corresponds to $b \geq \tfrac{1}{2}\ln[q/(q-2)]$, we fail to derive the explicit formula of spectral density and have to restrict ourselves with a graphical analysis. Figure 2a shows the curve of $b \geq \tfrac{1}{2}\ln[q/(q-2)]$ for a Bethe model at $q = 4$ and, for reference, the similar curve for a 2D Ising model. It is seen that the both curves almost coincide. Moreover, after computing the free energy for the two models, it is easy to show that in the close vicinity of critical points $\beta$ the difference between the magnitudes of free energy is no greater than 1.5%, the points themselves also being very close in magnitude. Though having this numerical resemblance, the two models demonstrate fairly different behavior. The reason is the big differences in the spectral densities, which are clearly seen in considering the derivatives of function $\Psi(E)$. Figures 2b and 2c show the first and the second derivatives of the spectral function for the Bethe model, the similar curves for the 2D Ising model being shown for comparison purposes. It is easy to notice that the derivatives behave quite differently. In the Bethe model at critical points $E = \pm \frac{1}{2} q/(q-2)$ the first derivative has a cusp which corresponds to a jump of the second derivative. Since the jump is finite, the thermal capacity also changes stepwise without divergence.

2.4. The mean-field approximation is very often used to describe spin systems. It is interesting to see what kind of spectral function $\Psi(E)$ this approach corresponds to. Using (7) – (8) or solving the Bragg-Williams equation for a $q$-neighbor hypercube model, we get

$$\Psi(E) = \ln 2 - \frac{1}{2}\left[(1-\bar{\varepsilon})\ln(1-\bar{\varepsilon}) + (1+\bar{\varepsilon})\ln(1+\bar{\varepsilon})\right], \qquad \bar{\varepsilon} = |2E/q|^{1/2} \qquad (17)$$

This formula holds true for all values of energy except the point $E = 0$, where the spectral function and its derivatives have to be redetermined and set to zero. Expression (17) resembles (10) and (16), yet the properties of the model with this kind of spectrum are dramatically different from the properties of the models corresponding to (10) and (16). For example, Figures 2 give graphs of the function $\Psi(E)$ corresponding to (17) and its derivatives for $q = 4$. It is seen from Figure 2a that the curve 3, which corresponds to the mean-field approximation is considerably different from the curve 1, which corresponds to the 2D Ising model. This difference is still more significant in Figures 2b and 2c: the first derivative experiences a jump at $E = 0$, the second derivative also turns into zero abruptly at $E = 0$. To avoid misunderstanding, we point out once more that the expression (17) does not describe the actual spectral density of some particular model. We give it here only to show that the mean-field approximation corresponds to the replacement of the actual spectral function by the expression (17) with ensuing consequences. For example, the use of the mean-field approximation in the 2D Ising model results in curves 1 being replaced by the curves 3 in Figure 2.

**3. Results and discussion**

First let us note that the central portion of the spectral density ($|E| \ll 1$) is defined by the expression:

$$D(E) = \frac{2^N}{\sigma}\sqrt{\frac{N}{2\pi}}\,\mathbf{exp}\!\left(-\frac{1}{2}N\frac{E^2}{\sigma^2}\right), \qquad \sigma^2 = \frac{1}{2N}\sum_{i,j=1}^{N} J_{ij}^2 \tag{18}$$

which is common to all possible models [11]. In particular, $\sigma^2 = qJ^2/2$ for all of the above-mentioned $q$-neighbors models. Indeed, the spectral density observed in the experiment appears as a narrow peak with a half-width of about $\sim \sigma/\sqrt{N}$, which can be described by the Gaussian (18). Moreover, in practice it is very difficult to find the difference between the experimental peak and curve (18) because the difference is of the $O(N^{-1})$ order. It is seen in the above graphs that the differences in magnitudes of $\Psi(E)$ for different models are not large and become noticeable only for great values of energy ($|E|\sim 1$) corresponding to the far wings of the energy spectrum. This means that the differences in properties of different spin models are largely caused by distinctions in the distribution tail areas.

From (7) – (8) it follows that the phase transition occurs when the derivative $d\Psi(E)/dE$ has either an inflection or a cusp at a particular critical value of energy $E = E_c$. If the second derivative turns to zero at the inflection point, the thermal capacity experiences the divergence. When the first derivative has a cusp, the thermal capacity experiences a jump of a finite magnitude. In the case of the 2D Ising model the examination of the curves shows that the nature of the phase transition is determined by the behavior of the spectral function $\Psi(E)$ within a very narrow range of energy. Indeed (see Fig. 1c), very narrow peaks (spectral regions $E = E_c \cdot (1 \pm 10^{-4})$ responsible for the phase transition) are clearly seen on the smooth curves of the function $d^2\Psi(E)/dE^2$. It is evident that even a negligibly small change in the spin interaction will result in either these peaks disappearing completely, or the second derivative failing to reach a zero value and, correspondingly, the logarithmic divergence of the thermal capacity at the critical point disappearing. It follows from the above that the logarithmic divergence of the thermal capacity predicted by Onsager is a unique fact that can be realized only when rigorous restrictions are applied to the energy spectrum.

From the above it also follows that one should handle the results obtained from the rough modelling of the spectral density very carefully. Indeed, slight variations of $\Psi(E)$ can lead to a radical change of vital parameters even in the case when $\Psi(E)$ is very similar to the spectral function of an actual system. Particularly, it is shown in paper [12] that the modelling of the spectral density of a 2D Ising model by using the so-called n-vicinity method allows for excellent numerical agreement with Onsager's results. On the other hand, there is a qualitative discrepancy (a jump of spontaneous magnetization at the critical point) that casts doubt on the results of the method used in [12].

In conclusion, we note that the models where $\Psi(E)$ is a piecewise smooth function have to be examined separately. We based our analysis on supposition that the function $\Psi(E)$ is a continuous and differentiable on the whole interval $E \in [E_0, -E_0]$. This supposition is rather reasonable from the physical point of view; however, from it follows that $\Psi(E)$ is a single-humped function. Indeed, the heat capacity $C = -\beta^2 d^2 f / d\beta^2$ being the dispersion of energy is positive according to the definition. Consequently, from the second of expressions (8) it follows that

$$d^2\Psi / dE^2 \leq 0. \tag{19}$$

With regard to this inequality, the function $\Psi(E)$ has only one maximum.

**Acknowledgments**. The work is partially supported by the RFBR grant #15-07-0486.

[1] R.J.Baxter. Exactly solved models in statistical mechanics. London: Academic Press (1982).

[2] H.Stanley. Introduction to phase transitions and critical phenomena. Clarendon Press, Oxford (1971).

[3] H.A.Kramers, G.H. Wannier. Phys. Rev. 60, 252–262 (1941).

[4] R. Kubo. An analytic method in statistical mechanics. Busserion Kenkyu 1 (1943) 1–13.

[5] M.Kac, J.Ward. Phys.Rev., 88, 1332 (1952).

[6] R.Becker, W.Doring. Ferromagnetism. Springer Verlag, Berlin, 1939.

[7] J.M. Dixon, J.A. Tuszynski, E.J. Carpenter. Physica A, 349, 487–510 (2005).

[8] B. V.Kryzhanovsky, L.B. Litinskii. Doklady Mathematics, 90, 784–787 (2014).

[9] L.Onsager. Phys.Rev., 65, 117 (1944).

[10] H.A.Bethe. Proc.Roy.Soc. (London), A150, 552 (1935)

[11] Ya. Karandashev, B. Kryzhanovsky. Opt. Memory & Neural Networks, 19, 110 (2010).

[12] B. Kryzhanovsky, L.Litinskii. Physica A, Vol. 468, pp. 493–507, 2017.